# Giant Oscillating Thermopower at Oxide Interfaces

*Ilaria Pallecchi[1], Francesca Telesio[1], Danfeng Li[2], Alexandre Fête[2], Stefano Gariglio[2], Jean-Marc Triscone[2], Alessio Filippetti[3], Pietro Delugas[4], Vincenzo Fiorentini[5], and Daniele Marré[1],\**

[1] CNR-SPIN and Genova University, Department of Physics, via Dodecaneso 33, 16146-Genova, Italy

[2] Departement of Quantum Matter Physics, University of Geneva, 24 Quai E.-Ansermet, 1211 Geneva 4, Switzerland

[3] CNR-IOM UOS Cagliari, c/o Dipartimento di Fisica, Università di Cagliari, S.P. Monserrato-Sestu Km.0,700, Monserrato (Ca), 09042, Italy.

[4] CompuNet, Istituto Italiano di Tecnologia - IIT, Via Morego 30, 16163 - Genova, Italy.

[5] Dipartimento di Fisica, Università di Cagliari, and CNR-IOM, S.P. Monserrato-Sestu Km.0,700, Monserrato (Ca), 09042, Italy.

**Corresponding Author**

* E-mail: marre@fisica.unige.it

**Understanding the nature of charge carriers at the LaAlO$_3$/SrTiO$_3$ interface is one of the major open issues in the full comprehension of the charge confinement phenomenon in oxide heterostructures. Here, we investigate thermopower to study the electronic structure** in LaAlO$_3$/SrTiO$_3$ at low temperature as a function of gate field. In particular, under large negative gate voltage, corresponding to the strongly depleted charge density regime, thermopower displays record-high negative values of the order of $10^4$-$10^5$ µV/K, oscillating at regular intervals as a function of the gate voltage. The huge thermopower magnitude can be attributed to the phonon-drag contribution, while the oscillations map the progressive depletion and the Fermi level descent across a dense array of localized states lying at the bottom of the Ti 3*d* conduction band. This study is the first direct evidence of a localized Anderson tail in the two-dimensional (2D) electron liquid at the LaAlO$_3$/SrTiO$_3$ interface.

**Introduction**

The electronic properties of the two-dimensional (2D) electron system found at the LaAlO$_3$/SrTiO$_3$ (LAO/STO) interface [1], characterized by Hall-measured sheet carrier densities $n_{2D}$~2-8·$10^{13}$ $cm^{-2}$ and a thickness of a few nm, are deeply affected by the electronic confinement which leads to an anisotropic spatial extension of Ti 3*d* conduction bands and to a subband structure of the $t_{2g}$ levels. This geometry favors the spontaneous charge localization within $d_{xy}$ levels spatially very confined ($\cong$ 2 nm) on the SrTiO$_3$ (STO) side of the interface. This electronic confinement and breaking of inversion symmetry is also at the origin of a strong Rashba spin-orbit coupling [2].

When the electronic density is large enough, Ti 3*d$_{xz}$*-*d$_{yz}$* subbands with a larger spatial extension are progressively filled with consequences on magnetotransport and possibly on superconductivity [3,4]. Less well understood is the nature of the low-density regime on the verge of localization, a key aspect for field-effect applications. Experiments have shown that a sharp transition to a highly resistive state occurs at a critical density of 0.5-1.5·$10^{13}$ cm$^{-2}$ [5]. Considerations based on the polarization catastrophe model [6] suggest that a large portion of the electron charge present at the interface may actually be trapped in localized levels located in energy somewhere below the band mobility edge. The presence of electrons localized by impurities, disorder, or polaronic behavior, has been reported in a number of works [7,8,9,10]. So far, however, only indirect evidences of these localized states were furnished, based on features in optical and transport properties.

Field-effect is unquestionably the most powerful approach to study the fundamental properties of oxide heterostructures at varying charge density. Under gate voltage, the phase diagram can be cleanly and reversibly explored without the complications inherent to chemical doping, which may easily result in additional effects blurring the intrinsic behavior of these systems. For the LAO/STO system, field-effect has been extensively used to reveal tunable superconductivity [11,12], strong Rashba spin-orbit [2,3], enhanced capacitance and negative compressibility [13], and magnetic effects [14,15,16,17]. Several works [5,13,18,19] in particular have emphasized the correlated/localized nature of the electron carriers at large negative gate voltage, corresponding to a strongly charge-depleted regime characterized by sheet resistance of the order MΩ or higher, and densities lower than $n_{2D}$~$10^{13}$ cm$^{-2}$.

Among transport measurements, the most direct probe of the electronic properties of a metal is certainly furnished by thermopower (or Seebeck coefficient, $S$). In contrast with magnetoresistivity (i.e. Shubnikov de Haas) measurements [20,21] which require high-mobility samples and strong magnetic fields, $S$ can be equally well measured at low and high temperature and for the whole range of Hall-measurable charge densities. Specifically, $S$ is the sum of two terms, the diffusive Seebeck ($S_d$), which is the potential drop generated by electrons diffused by a thermal gradient, and the phonon-drag Seebeck ($S_g$), i.e. the additional electron current generated by the coupling of electrons with diffused phonons. $S_d$, related to the energy derivative of the conductivity[22], is much less crucially dependent on the scattering regime than conductivity or mobility, and represents a very sensitive probe of the electronic density of states (DOS) of the system. On the other hand, $S_g$ is directly related to the electron-phonon coupling (EPC), and may be used as a probe for a quantitative evaluation of the momentum relaxation rate of the electrons due to EPC[23].

Note that very large values of the Seebeck coefficient have been measured in titanium oxides, even in the shape of three-dimensional samples (see for example ref. [24] and [25]). More recently, Ohta and coworkers [26,27,28] have studied the enhancement of the Seebeck effect driven by charge confinement in titanate heterostructures, both in the diffusion and drag regimes.

In this work, we perform Seebeck measurements under field effect, focusing on the depletion regime. We find that $S$ assumes huge negative values and oscillates while the Fermi level is progressively lowered. We consider this behavior as an evidence of charge localization below the bottom of the conduction band.

**Results**

Transport and thermopower under applied gate voltage of several LAO/STO samples are measured. Since all the samples show similar results, only one of them will be presented in the following (the others are described in the Supplementary Note 1 and shown in Supplementary Figure 1). Seebeck effect is measured in a home-made cryostat, from 4.2K to room temperature, using an ac technique [29], in the configuration sketched in Figure 1A). Hall effect and resistivity data are measured in a Quantum Design Physical Properties Measurement System (PPMS), from 2K to room temperature and in magnetic field up to 9 T. Both dc and ac techniques are used. In particular, in the depletion regime, the ac lock-in technique is preferable, in that it allows very low currents (~10 nA) to be applied and thus pinch-off problems to be avoided.

In Figure 1B), the Seebeck coefficient measured at 4.2 K as a function of the gate voltage $V_g$ is displayed, for two different thermal cycles. $S$ is negative at any voltage, indicating electron-like carriers. For positive $V_g$ (see inset) which scans the charge accumulation regime, $S$ shows the magnitude (~100 µV/K) typically seen in STO bulk or heterostructures [30,31,32] and varies by no more than 60% across the whole positive range, exhibiting a non-monotonic behavior. In this accumulation regime the measured magnitude of $S$ first displays a very small increment up to $V_g$ ~ 20 V, and then it decreases regularly at larger $V_g$; assuming diffusive regime, the lowering of $S$ is simply a consequence of the Fermi energy increase occurring along with the progressive charge accumulation. If one looks now at the negative $V_g$ region, after a featureless gate voltage interval of ~7 V, something really astonishing happens: $S$ progressively bursts to record-high values of the order of ~$10^4$ µV/K (it changes from $S$= -80 µV/K at $V_g$ = -5 V to -26000 µV/K, that is more than two orders of magnitude, for a $V_g$ variation of only 10 V). Furthermore, $S$ dramatically oscillates with $V_g$. These oscillations are regularly distributed, and roughly occur at voltage intervals of 0.6 V. Above a maximum limit of $V_g$~ -15 V, no thermoelectric voltage signal can be detected experimentally anymore, the sample entering a highly insulating state. We may argue, however, that in principle the exponential rise of $S$ to even greater negative values and the associated oscillations could be continued up to the ideal limit of complete charge depletion (in the Supplementary Note 1, a sample with $S$ reaching ~ -$10^5$ µV/K is shown as well). It is interesting to note that slight shifts of the $S$ curves are observed after different cycles of temperature and $V_g$, as explained in the Methods and Supplementary Note 1 sections, however the oscillating and diverging behavior is always present. In particular, oscillating and diverging $S$ curves are systematically observed at low negative $V_g$ values after a "forming" protocol of $V_g$ cycle at low temperature. The protocol is to apply at low temperature first the highest positive voltage $V_{max}$ to be used and then to sweep the voltage between $V_{max}$ and $V_{min}$. This procedure allows reproducible measurements to be obtained.

In Figures 2A) and 2B) the characterization of electric and thermoelectric properties as a function of $T$ (at $V_g$=0) and $V_g$ (at $T$=4.2 K) is reported, respectively. From the measurement of longitudinal and Hall resistances, sheet carrier density $n_{2D}$ (which as usual we identify with the inverse Hall resistance assuming the Hall factor equal to unity and single band description) and Hall mobility $\mu$ are extracted. As a function of $T$ and at zero gate voltage, the sample shows features typical of standard LAO/STO interfaces [30]: $S$ exhibits a phonon drag peak at low temperature and is linear above 50K, as expected in the diffusive regime. $R_{sheet}$ decreases with decreasing $T$, as expected for phonon-limited metallic behavior, and saturates at a residual value of 700 $\Omega$. The sheet carrier density $n_{2D}$ extracted from the inverse Hall constant varies, depending on the samples, in the range 3-8·$10^{13}$ cm$^{-2}$ as is common for LAO/STO samples. A slight $R_H$ downturn at low temperature has also been observed in other similar samples [33] and may be interpreted in terms of charge delocalization related to the increased screening at low temperature, where the dielectric permittivity is very large. Mobility has a low-$T$ value around 380 cm$^2$V$^{-1}$s$^{-1}$ and decreases approximately as $T^2$ at high $T$, as expected from scattering by optical phonons. The same properties measured at fixed $T$=4.2 K and varying $V_g$ (Figure 2B)) depict an accumulation-depletion scenario, with $R_{sheet}$ and $n_{2D}$ being decreasing and increasing functions of $V_g$, respectively. $R_{sheet}$ shows a sharp change in slope at $V_g$=0, that is at the border of accumulation and depletion regions, similarly to what was reported in previous field-effect measurements at low $T$ [12,18], and reaches values of the order of $10^4$-$10^5$ $\Omega$ for $V_g <$ - 40 V corresponding to a maximally depleted sheet density $n_{2D} \sim 8 \cdot 10^{12}$ cm$^{-2}$. Below this threshold, $R_{sheet}$ cannot be measured any longer since a sharp metal-insulator transition (MIT) occurs. Consistently, we find the Hall mobility to decrease with negative $V_g$ up to the lowest measurable value, in agreement with literature [19].

**Discussion**

From the measurements two essential concepts emerge: i) on the positive side of the gate field, transport properties are compatible with ordinary 2D electron gas (2DEG) behavior in the charge accumulation regime; a switch towards negative gate values drags the system towards a highly depleted regime characterized by a sharp rise of resistivity and decrease of mobility, prefiguring a transition from metallic to localized charge transport behavior; ii) At low $T$ this progressive depletion is indicated by rapidly diverging and dramatically oscillating Seebeck coefficient. In the absence of any magnetic field, this huge oscillating thermopower has no exact precedent in literature, to our knowledge. In a small number of works a large field effect enhancement of the Seebeck coefficient is reported. The closest analogies with our result are probably the diverging thermopower and electrical resistivity measured in Silicon 2DEG [34], occurring near the metal-insulator transition and attributed to Anderson localization, and the oscillating thermopower as a function of the carrier concentration found in GaAs/AlGaAs 2DEGs [35] and in single-layer MoS$_2$ [36], attributed respectively to electron-localization phenomena and variable range hopping mechanism. In both cases, however, the measured $S$ coefficients ($\sim 10^3$ μV/K for Si and $\sim 10^2$ μV/K for GaAs 2DEG) remain well below our values, and in both papers the dominant thermoelectric contribution is attributed to electron diffusion. While we agree that electron localization must be the appropriate electronic structure landscape for this phenomenon, our standing point is that the measured thermopower is incompatible with electron diffusion, and must be primarily due to phonon-drag, which is known to dominate over diffusion in certain ranges of temperature and densities where the EPC is strong.

Evidence of large EPC in STO bulk and STO-based heterostructures related to electron localization has been reported in several works [7,8,37,38]. EPC can ignite large phonon-drag according to the following expression for $S_g$ in the $j$ direction (the detailed formulation [39,40,41] is reported in the Supplementary Notes 4-7):

$$S_j^{\mathrm{g}} \approx \left( \frac{1}{\sigma_j T^2} \right) \sum_q \left( \frac{\hbar \omega_q}{\tau_{\mathrm{ph}}^{-1}(\mathbf{q}) + \tau_{\mathrm{ep}}^{-1}(\mathbf{q})} \right) \sum_{n\mathbf{k}, n\mathbf{k'}} \Gamma_{n\mathbf{k}, n\mathbf{k'}}(\mathbf{q}) V_j(n\mathbf{k}, n\mathbf{k'}, \mathbf{q}) \qquad (1)$$

where $\sigma_j$ is the electron conductivity, $V_j$ a velocity factor, $\Gamma_{n\mathbf{k},n\mathbf{k'}}$ the electron-phonon scattering matrix, $\tau_{\mathrm{ep}}$ the phonon relaxation time due to EPC, while $\tau_{\mathrm{ph}}$ sums up all the other relevant relaxation mechanisms (e.g. phonon-phonon, phonon-boundary, phonon-impurity scatterings), and the sum is over all the relevant phonon vectors ($\mathbf{q}$) and modes, as well as electronic bands ($n$) and wavevectors ($\mathbf{k}$). In standard metallic (accumulation) regime, $S_g$ is only relevant in the narrow $T$ interval (typically around $T \sim \theta_D/5$, $\theta_D$ is the Debye temperature) where $T$ is large enough to activate a substantial number of phonons, but not so large as to make phonon-phonon scattering ($\tau_{\mathrm{ph}}^{-1} \sim T^3$) dominant over EPC, since for $\tau_{\mathrm{ph}}^{-1} >> \tau_{\mathrm{ep}}^{-1}$ $S_g$ quickly vanishes.

This scenario is dramatically altered in case of strong depletion. First, electron localization largely suppresses conductivity, thus favoring large $S_g$, according to Equation (1); at variance, $S_d$ has a much weaker dependence on $\sigma_j$ (see Supplementary Equation (32) and the related discussion in the Supplementary Note 6), and cannot be amplified as much by a decrease of conductivity. Second, EPC may become very large in the so-called "dirty limit", that is for electron diffusion lengths $L$ much shorter than phonon wavelengths ($qL<<1$) [23,34]. In particular, this condition holds for narrow two-dimensional electron systems [40] in the confined direction (e.g. $z$), since strong EPC occurs for any $q_z<<1/t$, ($t$ being the electron liquid thickness) thus any phonon wavevector contributes, eventually up to the Debye frequency. In other words, in the limit $t\rightarrow0$ EPC may be large since even very short-wavelength acoustic phonons can couple to electrons. A further burst to EPC may come from the polar character of the system, which can lead to a large coupling between electrons and the electric field associated with acoustic phonons of very-long wavelength. The peculiar characteristics of 2-dimensional electron liquid (2DEL) in LAO/STO under large negative gate voltage are thus highly favorable to the occurrence of strong EPC.

As a quantitative assessment of our analysis, we calculate the Seebeck coefficient and resistivity for a multiband effective-mass model (details are given in the Supplementary Notes 4-8 ). The model includes a single delocalized conduction band of $t_{2g}$ $d_{xy}$ symmetry, which is sufficient to describe the low-density regime $n_{2D} <10^{13}$ cm$^{-2}$ according to First-Principles calculations 4 . Below the conduction band bottom, we insert (see Figures 3A), 3B), and 3C)) a series of low-density, low-mobility polaronic states (for brevity called "localized" hereafter), with the intent of mimicking a disorder-induced Anderson tail, consistent with the scenario proposed in previous works [7,8,10]. Our model is also consistent with a scenario dominated by Mott localization, often invoked to explain other correlation phenomena observed in this system (see e.g. capacitance enhancement described in Ref.13), since electrons can localize on Ti$^{3+}$ sites at very low charge density, eventually helped by structural deformations (i.e. polarons), even in absence of disorder. Since the measured Seebeck coefficient shows 12 oscillations in the range between -7V and -14V, and the Hall measurement indicates in this voltage range a modulation of the carrier density of $n_{2D} \sim 6 \times 10^{11}$ cm$^{-2}$, in the model we include 12 localized states, whose integrated DOS amounts to $n_{2D} \sim 6 \times 10^{11}$ cm$^{-2}$. The basic characteristics of these states (energies, effective masses, bandwidth) are set to qualitatively reproduce the measured Seebeck in terms of oscillation frequency and amplitudes, as well as the Seebeck absolute value. The model is finalized to reproduce the following scenario: at zero voltage, the lowest conduction band is occupied by a mobile carrier density ($n_{2D} \sim 1.2 \times 10^{13}$ cm$^{-2}$) similar to that usually found in LAO/STO. Then, it is assumed that the substantial effect of a negative $V_g$ is the progressive charge depletion, thus the progressive lowering of $E_F$, so that mapping $S(E_F)$ is equivalent to mapping $S(V_g)$ (additional details are given in the Supplementary Note 8 ).

In Figure 3, the calculated $S(E_F)$ and $\rho(E_F)$ reproduce qualitatively the most important features observed at varying $V_g$. In the experiment, negative $V_g$ causes mild changes in $R_{sheet}$ and $S$ up to a certain threshold, which we can dub "regular behavior". In our interpretation, this threshold corresponds to the point where $E_F$ reaches the Conduction Band Bottom (CBB) edge (zero energy in the Figure 3d-3f). For $E_F \leq E_{CBB}$ a dramatic upturn begins: $E_F$ enters the region of localized states, $\rho$ rises to $\Omega \cdot$cm values and $S_g$ (negligibly small in accumulation) quickly bursts up to $10^5$ µV/K in magnitude. During the $E_F$ descent through the series of increasingly localized states, $S_g$ oscillates in correspondence with the crossing (depletion) of each state. $S_d$ also oscillates in correspondence with localized states, but with several fundamental differences from $S_g$: a) the magnitude of $S_d$ ($\sim 10^2$ µV/K) is in the usual range for LAO/STO (or even bulk SrTiO$_3$) at low $T$, while $S_g$ is 3-4 orders of magnitude larger; b) $S_g$ increases rapidly in magnitude with the lowering of $E_F$, as a consequence of the $1/\sigma$ dependence, while the oscillation-averaged $S_d$ changes very smoothly with $E_F$; c) the amplitude of $S_d$ oscillations is comparable with the averaged $S_d$ value itself. Points b) and c) make $S_d$ incompatible with the monotonic descent observed in the experiment. These aspects (discussed in detail in the Supplementary Note 8 ) are derived from the fundamental characteristics of $S_d$, not from the specific approximation used in the model, thus they represent sound arguments to rule out diffusive Seebeck as the main contribution to the values measured in the strongly depleted regime. We remark that other forms of transport typical of localized carriers (variable-range hopping and thermal activated hopping) were considered as a possible source of the observed oscillations, but they were found incompatible with either the huge oscillating Seebeck or the resistivity values.

Notice that in the simulation $\rho$ also shows oscillations, which are not observed experimentally. A non-oscillating behavior of $R_{sheet}(V_g)$ in presence of oscillations in $S(V_g)$ has been previously noticed in 2DEGs (GaAs/AlGaAs [42,43] and MoS$_2$ [44]). We believe that this is a consequence of the different nature of the two measurements (open- vs. closed-circuit for $S$ and $R$, respectively) and the non-ohmic behavior of resistivity in the strongly depleted regime [45] (a more extended discussion can be found in the Supplementary Note 2).

In conclusion, we measure field-effect transport and thermoelectric properties of the 2DEL at the LAO/STO interface in a wide range of gate field values. Seebeck measurement is an exceptionally sensitive tool for the exploration of highly confined carriers in oxide heterostructures, since it displays a clear signal even for electronic states characterized by extremely low charge density and mobility, and capable to resolve electronic structures in the lower than meV range at very low temperature. In this respect, thermopower seems to be a far more effective probe than any photoemission techniques nowadays available (see Supplementary Note 3).

At low-$T$, for negative gate voltage larger in magnitude than a given threshold, the measured thermopower displays a diverging, oscillating behavior with negative values that are unprecedented in literature. With the help of model results, we interpret this finding as being due to the regime change from normal-metal to localized electron behavior, corresponding to the $E_F$ descent below the conduction band, and across a series of lower lying localized electronic states, characterized by large resistivity, large EPC, and in turn huge measurable phonon-drag.

This huge-Seebeck transient phase can be considered incipient of the insulating regime: it lives in a narrow charge density interval between the band-like metallic density ($n_{2D} \geq 10^{13}$ cm$^{-2}$) and the insulating phase ($n_{2D} \leq 8$-$9$ $10^{12}$ cm$^{-2}$), so that only a handful of these states (recognizable by the number of $S$ oscillations) are visible. If the polarization catastrophe scenario is correct, we expect that a further lowering of $E_F$ would bring us right into the insulating phase where a large quantity of localized charge ($n_{2D} \sim 10^{14}$ cm$^{-2}$) is present. We argue that this charge is associated to diverging S values and conductivity too low to be detected by any means. Thus, the thermopower oscillations measured in this work are a strong, direct signature of the presence of localized states in the LAO/STO system.

**Methods**

**Sample preparation and measurement setup**

The samples are prepared as described in detail in reference [46]. Pulsed laser deposition is used to grow an epitaxial layer of LaAlO$_3$ (4 to 10 unit cells thick) on a (001)-oriented TiO$_2$-terminated SrTiO$_3$ substrate. The substrate temperature is about 800°C and the oxygen pressure $10^{-4}$ Torr. The laser fluence is estimated at 0.6 J/cm$^2$. Reflection high energy electron diffraction (RHEED) intensity oscillations and patterns show layer-by-layer growth and good crystalline quality. The growth rate is around 60 laser shots/unit cell. After deposition, the oxygen pressure is raised to 200 mbar and the sample is kept at 500°C for one hour before being cooled down to room temperature. X-ray diffraction patterns confirm that the LaAlO$_3$ layers are epitaxial and crystalline while atomic force microscope (AFM) images show atomically flat terraces on the LaAlO$_3$ surface, with terrace-edge heights corresponding to one layer of LaAlO$_3$. Other details about sample preparation are found in references [47,48,49]

The samples measured in this work behave in a very similar way, so that data on only one of them are presented. One sample is patterned in the shape of a Hall bar (channel width 1.5 mm) for better definition of geometrical factors and is used as a reference to confirm the exact relationship between measured properties. The measurement configuration sketched in Figure 1A) shows that the *ac* heat flows along the 010 direction (the longer side) of the 10 mm X 5 mm crystal. The typical power fed to the ac heater is ~1 mW. The voltage contact separation, equal to thermocouple separation, is 1.5-3 mm. The back gate electrode is a 3 mm X 5 mm gold pad evaporated on the backside of the 0.5 mm thick substrate. The leakage current to the gate electrode is always monitored to be below 10 nA during the *S* measurements.

The Seebeck coefficient is measured in a home-made cryostat, from 4.2K to room temperature, using an ac technique [29]. The period of the sinusoidal power supplied to the sample is 150 s and the applied thermal gradient is around 0.3K across a distance of ~2 mm. Hall effect and resistivity data are measured in a Quantum Design PPMS system, from 2K to room temperature and in magnetic field up to 9 T. Both dc and ac techniques are used. In particular, in the depletion regime, the ac lock-in technique is preferable, in that it allows very low currents (~10nA) to be applied and pinch-off problems to be avoided.

The electrical and thermoelectrical responses to field effect of all the samples exhibit the general features presented in this work, in particular the divergent and oscillating S in the depletion regime. Nevertheless, some specific details (threshold voltage for the occurrence of the depletion regime, as well as exact magnitude of $S$) may vary from sample to sample and from thermal cycle to thermal cycle. We believe that this behavior originates from the fact that the electrostatic landscape around the two-dimensional electron liquid may change significantly, due to empty and filled traps within the $SrTiO_3$ substrate. This may affect the screening of the gate electric field and thus may be the primary source of changing threshold voltage and details of experimental curves. On the other hand, reproducible $S(V_g)$ curves are obtained after a "forming" protocol of $V_g$ cycle at low temperature **Errore. Il segnalibro non è definito.**, carried out by applying, at low temperature, first the highest positive voltage $V_{max}$ (usually ~200V) to be used and then sweeping the voltage between $V_{max}$ and $V_{min}$. This procedure allows reproducible measurements to be obtained, even sweeping $V_g$ back and forth, as long as the sample is maintained at low temperature. The stability and reproducibility of Seebeck measurements after the forming protocol is also checked by measuring Seebeck signal for hours at fixed temperature and gate voltage, which yields constant curves.

In any case, considering all the different sets of measurements, it is clear that all the measurements present the same striking common features, that is evenly spaced oscillations as a function of gate voltage and diverging Seebeck coefficient in the tens of mV/K range in the depletion regime, which support the general character of our results and their interpretation.

We point out that there is a clear discrepancy between, on the one hand, the low T limit values in electrical and thermo-electrical transport curves in Figure 2A) and, on the other hand, the zero gate voltage values of thermo-electrical and electrical transport properties in Figures 1B) and 2B), respectively. Indeed, the $S$ value at the lowest temperature in Figure 2A) is around -240µV/K slightly different from the $V_g=0$ value of Figure 1B). Similarly, the $n_{2D}(V_g=0)$ value in Figure 2B) is much smaller than the low temperature value displayed in the third panel of Figure 2A). This is due to the fact that the electron state at the interface set up by the forming protocol is stable only at the lowest temperatures. Data in Figures 1B) and 2B) are measured after the forming protocol, while in Figure 2A), data are measured before the forming protocol.

**Theoretical modeling**

Diffusive thermopower and electric resistivity is calculated through the Boltzmann Equation in the relaxation time approximation [50]. The electronic structure is simulated using a variant of the multi-band effective mass modeling previously employed to describe STO/LAO [30], and the electronic relaxation time is modeled in terms of common energy-dependent analytic formulae (the Brooks-Herring formula for impurity scattering and the deformation potential approach for acoustic phonon scattering). Phonon drag formulation is based on the coupled Boltzmann Equations for electrons and phonons, in relaxation time approximation [39,40]. Since the low-temperature limit is required, only the electron-acoustic phonon scattering is included in the phonon-drag. Acoustic phonons are modeled according to the deformation potential approach, plus a piezoelectric scattering to account for the polar character of STO. The theoretical model is built to reproduce n-doped STO bulk, even with the presence of localized states below the conduction band (see Supplementary Figure 2 and 3). LAO/STO quantum well DOS is modeled with 12 equally spaced localized levels lying below the conduction band bottom. In Supplementary Figure 4 the individual contributions of each electronic state to phonon-drag, diffusive Seebeck, and conductivity are shown, while main features of each level in terms of effective mass and electronic bandwidth are reported in Supplementary Table 1. Finally, the electron-phonon scattering rate at different doping level is reported in Supplementary Figure 5. The formulation is displayed in full detail in the Supplementary Note 4.

Acknowledgements

We acknowledge fruitful discussions with C.Aruta, A.Braggio, R.Buzio, C,Cancellieri, A.Gerbi, P.Orgiani and M.Salluzzo. This work was supported by: Italian MIUR through projects FIRB RBAP115AYN "Oxides at the nanoscale: multifunctionality and applications" and PRIN 2010NR4MXA "OXIDE", Università di Genova, Swiss National Science Foundation through the National Center of Competence in Research, Materials with Novel Electronic Properties, 'MaNEP', the "Thermoelectric oxides TEO" project and Division II and the European Research Council under the European Union's Seventh Framework Programme (FP7/2007-2013)/ERC Grant Agreement No. 319286 (Q-MAC).

P.D. acknowledges financial support by CompuNet, Italian Institute of Technology (IIT) under "Platform Computation". V.F. acknowledges support by CAR of Università di Cagliari. P.D, A.F. and V.F. acknowledges financial support by Fondazione Banco di Sardegna Projects, and computational support by CINECA (Casalecchio di Reno, Italy) and CRS4 (Pula, Italy).

————————————————

**Author Contributions**

Samples were deposited by D.L. and A.F. Transport and thermoelectric measurements were performed by I.P., F.T. and D.M. Theoretical calculations and modelling were carried out by A.F. P.D. and V.F. All the authors contributed to data interpretation and to the manuscript preparation and approved the final version of the manuscript.

**Competing interests statement**

The authors declare no competing financial interests.

**Figure captions**

**Figure 1. Seebeck measurement configuration and behavior under gate field of a LAO/STO interface A)** Sketch of the sample and experimental configuration for the Seebeck measurements: the two dimensional electron liquid lies in the (001) plane of the $SrTiO_3$ while the thermal gradient is applied along the 010 direction. **B)** Seebeck coefficient versus gate voltage measured in a LAO/STO interface at 4.2 K. In the main panel, the different traces correspond to different thermal and $V_g$ cycles: the red curve is measured using an *ac* heat flow whose power is 3.5 times smaller than the one (~mW) used for the blue and green curves. The blue and red curves are measured with decreasing gate voltage, while the green curve is measured with increasing gate voltage. The measurement limit related to the finite input impedance of the instruments used for the measurement of the voltage is also indicated. In the inset, a blow-up of the accumulation regime ($V_g > 0$) is shown.

**Figure 2: Electric and thermoelectric properties of LAO/STO interfaces A)** Electric and thermoelectric properties of the two-dimensional electron liquid as a function of temperature. From top to bottom are Seebeck coefficient, sheet resistance, inverse Hall constant and Hall mobility. **B):** Electric transport properties of the LAO/STO interface at 4.2K as a function of the gate voltage, namely sheet resistance, inverse Hall constant and carrier Hall mobility.

**Figure 3: Electronic band structure of the two-dimensional electron gas (2DEG) emerging from the experimental results, and calculated transport properties A):** Sketch of the model band structure purposely built to reproduce the experimental results. Gray areas indicate valence and conduction states; the colored lines below the conduction states represent a tail of localized states. **B):** Actual Density of States (DOS) of the model band structure considered for the calculations. The shaded gray area is the

DOS relative to the Conduction Band Bottom (CBB) of $t_{2g}$ $d_{xy}$ orbital character. Below the CBB lies a tail of 12 localized states, placed at regular intervals of 3 meV from each other, indicated by different colors and type of lines. From the bottom: red solid, dotted, dashed, dot-dashed, and then the same sequence repeated in green and blue. Zero energy is fixed at the CBB. C): Integrated DOS per unit area. The DOS is normalized to obtain for the total charge density hosted by the 12 localized states, $n_{2D}$= 6x10$^{11}$ cm$^{-2}$ , that is the Hall-measured charge depleted by field-effect in the interval $V_g$=-14 V, -7 V where the huge Seebeck oscillations are visible. D): Phonon-drag calculated for the model DOS. The dotted vertical lines indicate the bottom energy of each localized state, the solid line is the CBB. $S_g$ oscillates at each intersection of $E_F$ with the bottom energies. E): Diffusive Seebeck; like $S_g$ it oscillates in correspondence with the depletion of each localized state, but in absolute value is about 3 orders of magnitude smaller than $S_g$. F): Electric resistivity $\rho$ in 3D. To obtain the sheet resistance, $\rho$ must be rescaled by the 2DEG thickness t, i.e. $\rho = R_{sheet} \cdot t$.

**Figure 1**

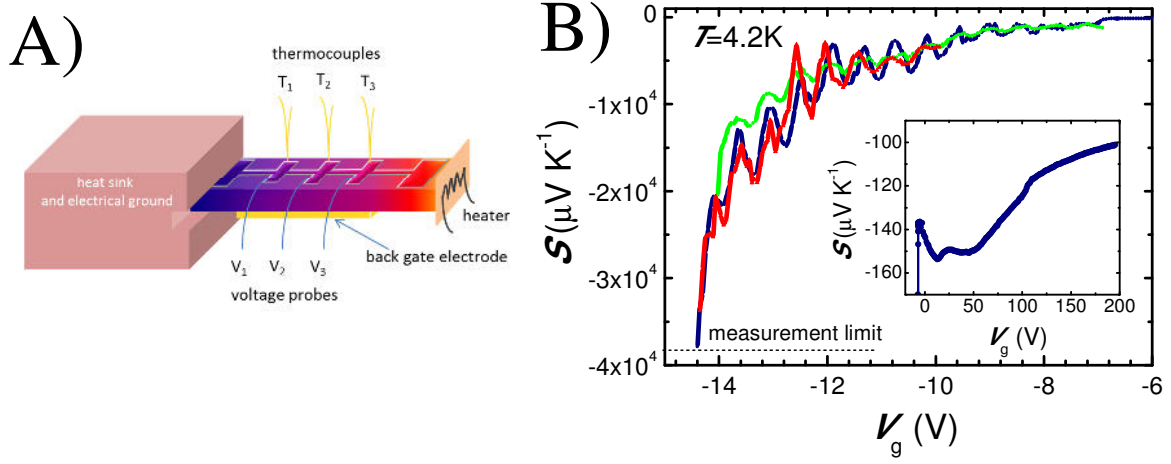

**Figure 2**

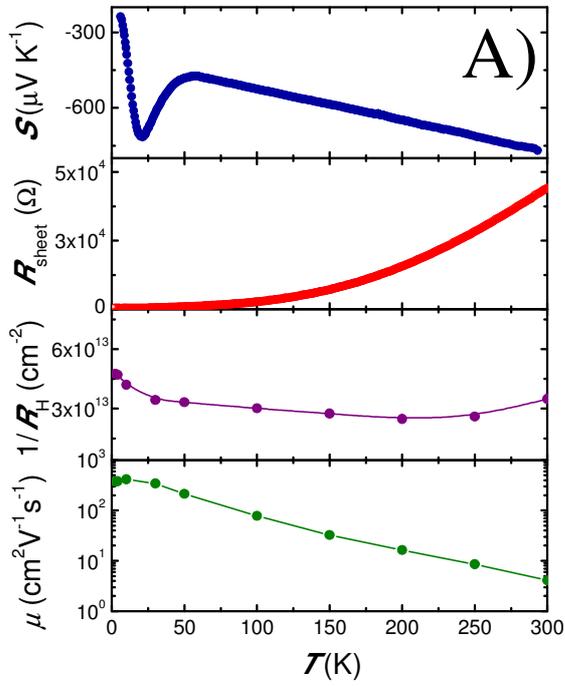 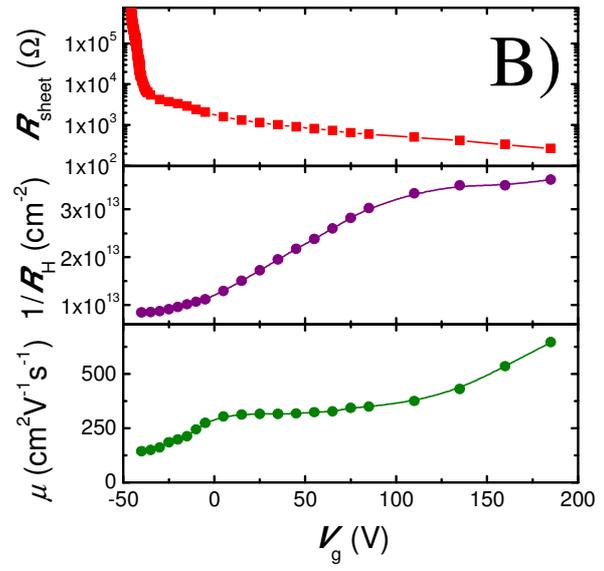

**Figure 3**

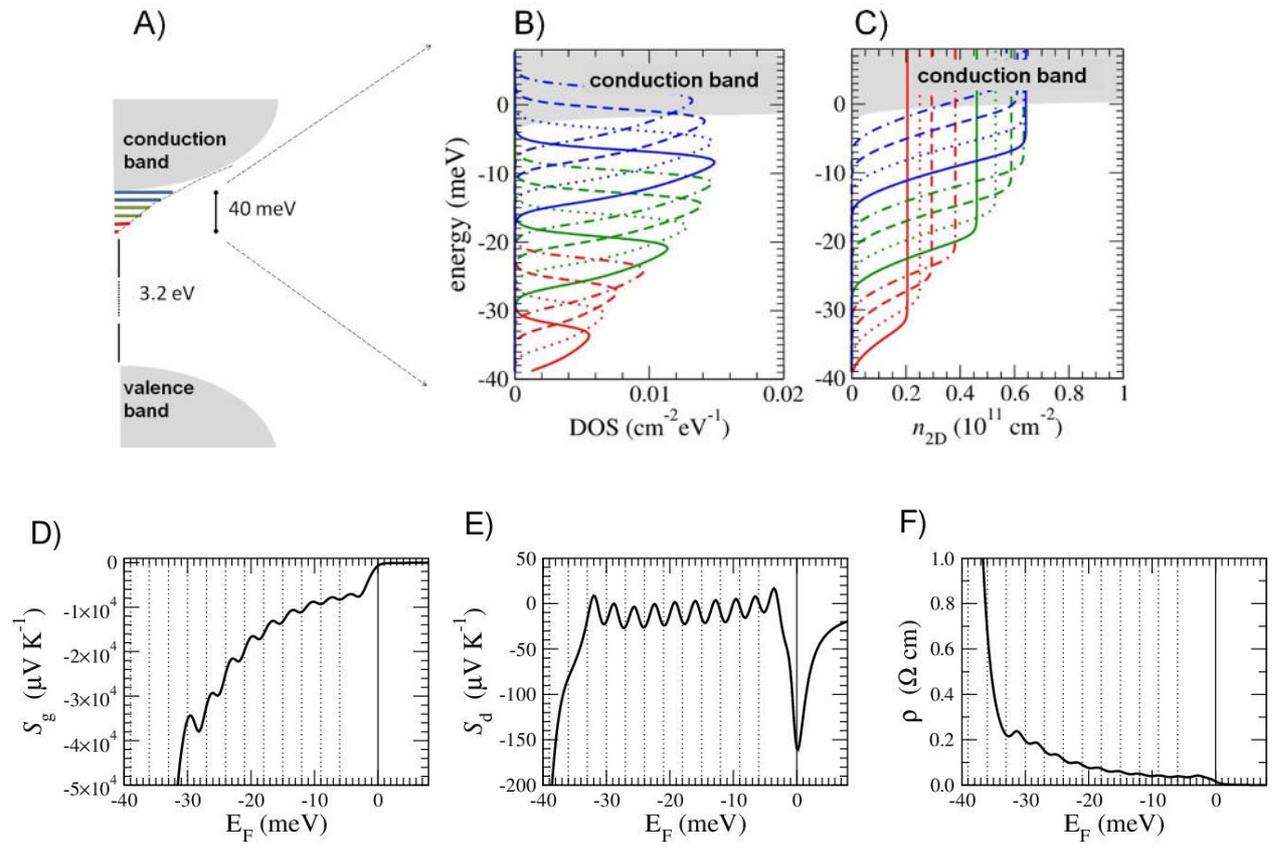

# Supplementary Information

# Giant Oscillating Thermopower at Oxide Interfaces

*Ilaria Pallecchi, Francesca Telesio, Danfeng Li, Alexandre Fête, Stefano Gariglio, Jean-Marc Triscone, Alessio Filippetti, Pietro Delugas, Vincenzo Fiorentini, and Daniele Marré* [*]

**SUPPLEMENTARY FIGURES**

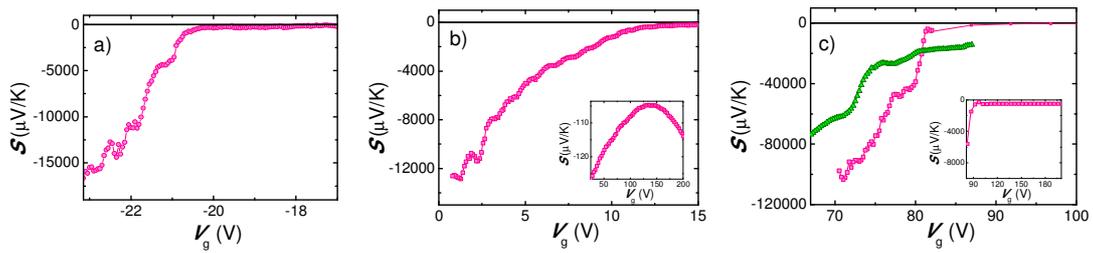

**Supplementary Figure 1:** **Seebeck coefficient versus gate voltage measured at 4.2K** in other LAO/STO interfaces, whose parameters are listed in the text.

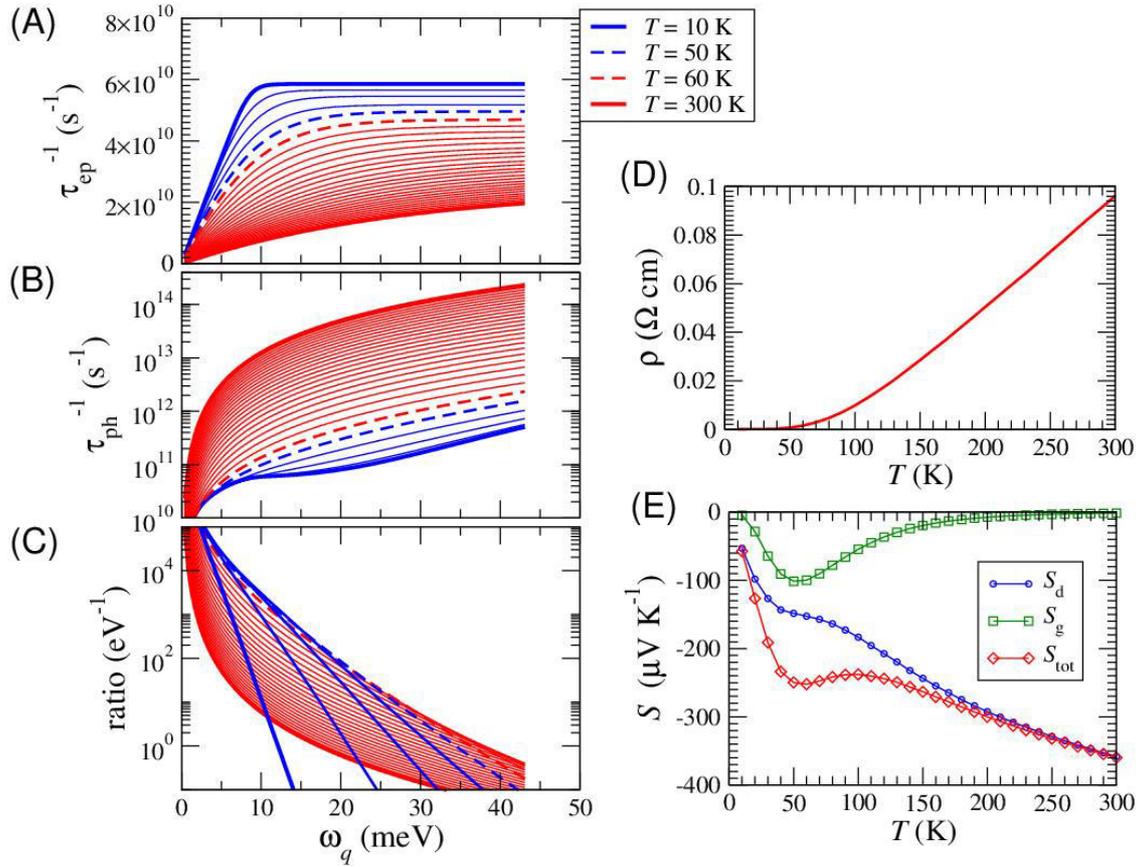

**Supplementary Figure 2: Calculations for n-doped STO bulk ($n_{3D}$= 2×10$^{19}$ cm$^{-3}$) represented by three $t_{2g}$ conduction bands** (A): electron-phonon scattering frequency as a function of phonon energy calculated from Equation (30); different curves span a temperature range from 10 K to 300 K in steps of 10 K. Blue to red color change highlights the regime change from growing to decreasing phonon-drag. (B): total phonon scattering frequency calculated adding the electron-phonon to the other contributions given in Equation (31). (C): relative scattering ratio (Equation (38)): it grows with $T$ up to 50-60 K (highlighted by the dashed lines), and then it starts to fall and decreases smoothly up to vanishing at room $T$. (D): DC resistivity, signaling metallic behavior. (E): total Seebeck and decomposition in diffusive and phonon-drag contributions.

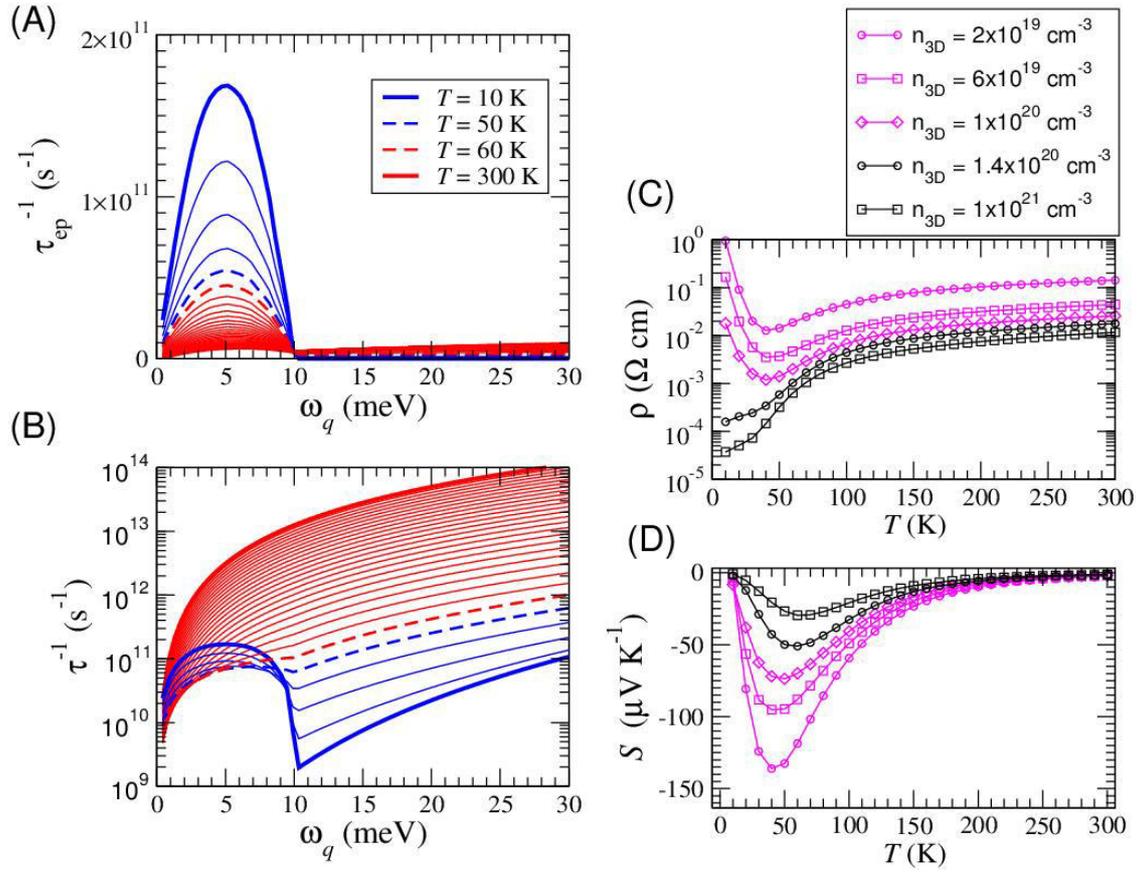

**Supplementary Figure 3: Calculations for n-doped STO bulk whose electronic structure consists in three $t_{2g}$ conduction bands and one localized state lying 10 meV below the CBB** (see text) (A): electron-phonon scattering frequency as a function of phonon energy calculated from Equation (30) at $n_{3D}$= 2×10$^{19}$ cm$^{-3}$; different curves span a temperature range from 10 K to 300 K in steps of 10 K. (B): total phonon scattering frequency at $n_{3D}$= 2×10$^{19}$ cm$^{-3}$ calculated as the sum of electron-phonon scattering in Equation (28) and other phonon scatterings given in Equation (29). (C): resistivity calculated at different doping concentrations (E): phonon-drag calculated at different doping concentrations

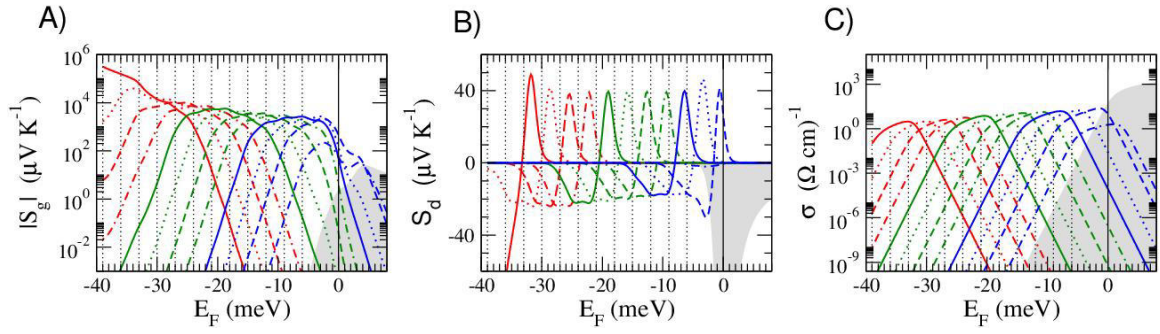

**Supplementary Figure 4: Individual contributions of each electronic state to phonon-drag, diffusive Seebeck, and conductivity, calculated for the model DOS in Figure 3 A):** Phonon-drag amplitudes due to individual electronic states. The dotted vertical lines indicate the bottom energy of each localized state, the solid line is the CBB. Line colors and styles relate each contribution to the corresponding state DOS in 3B). The gray-shaded area is the contribution of the lowest conduction band (gray-shaded area). **B):** Contributions to $S_d$ from each state. In contrast with $S_g$, the contributions due to the localized states (except for the conduction band) always cross the zero. C): State-by-state contributions to 3D conductivity; $\sigma_{sheet} = \sigma t$.

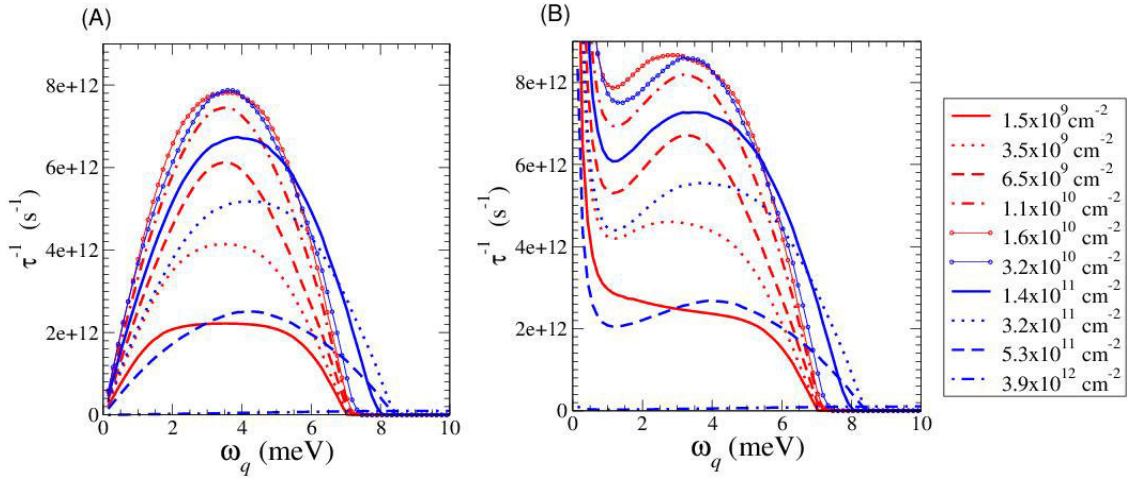

**Supplementary Figure 5: Electron-phonon scattering rate calculated from Equation (30) at fixed *T*=4.2 K and varying doping concentration** mimicking the progressive charge depletion for increasingly negative gate voltage. Different curves are for different Fermi energies (i.e. doping), spanning the range of localized states illustrated in Figure 3 of the main article. The red solid curve is for the lowest $E_F$; starting from low doping, the scattering first increases with $E_F$ (red curves) and then it decrease (blue curves). (A): only the deformation-potential contribution is included in the electron-phonon scattering (first term of Equation (22)). (B) both deformation potential plus piezoelectric scattering are included, according to Equation (22).

# SUPPLEMENTARY TABLES

**Supplementary Table 1: Bottom energy ($\varepsilon_b$), effective masses, and bandwidth ($W$) of the energy levels used to model the n-doped LAO/STO interface under large negative gate voltage**. The conduction band bottom (CBB) is placed at zero. Below CBB, a series of 12 localized states located at 3 meV from each other span a range of 40 meV. Effective masses increase and bandwidths decrease while going from higher to lower energies.

| $\varepsilon_b$ (meV) | $m_i^*$ ($m_e$) | $W$ (meV) |
|---|---|---|
| 0 | 0.7, 0.7, 8.0 | 1000 |
| -6 | 6, 6, 20 | 8.4 |
| -9 | 7, 7, 20 | 8.4 |
| -12 | 8, 8, 20 | 8.4 |
| -15 | 9, 9, 20 | 8.4 |
| -18 | 10, 10, 20 | 8.4 |
| -21 | 11, 11, 20 | 8.2 |
| -24 | 12, 12, 20 | 8.0 |
| -27 | 13, 13, 20 | 7.8 |
| -30 | 14, 14, 20 | 7.6 |
| -33 | 15, 15, 20 | 7.4 |
| -36 | 16, 16, 20 | 7.2 |
| -39 | 17, 17, 20 | 7.0 |

**SUPPLEMENTARY DISCUSSION**

**Supplementary note 1: Remark on the reproducibility of diverging and oscillating behavior of S in the depletion regime in different LAO/STO samples.**

Here below we show other $S(Vg)$ measurements carried out on different LAO/STO interfaces. The main features discussed in this work – i.e. diverging and oscillating behavior of S in the depletion regime and change of slope in electrical and thermo-electrical transport curves at the crossover between accumulation and depletion regimes - are present in all cases. In panel a) of Supplementary Figure 1, the measurement is carried out on a sample having similar sheet carrier density and sheet resistance as that presented in the main text (namely the regime of $S$ oscillations corresponds to $|S|\approx15$ mV/K, $R_{sheet}\approx10$ KΩ and $n_{2D}\approx6\times10^{12}$ cm$^{-2}$, while in the accumulation regime $|S|\approx 200$ μV/K, $R_{sheet} \approx 530$ Ω and $n_{2D}\approx2.7\times10^{13}$ cm$^{-2}$). In panel b), it is presented the behavior of a LAO/STO interface, where the regime of $S$ oscillations corresponds to $|S|\approx13$ mV/K and $R_{sheet}\approx100$ KΩ, while in the accumulation regime $|S|\approx100$ μV/K, $R_{sheet}\approx1$ KΩ and $n_{2D}\approx3\times10^{13}$ cm$^{-2}$. Finally, in panel c), it is presented the behavior of a different LAO/STO interface, where the regime of $S$ oscillations corresponds to $|S|\approx50$-$100$ mV/K and $R_{sheet}\approx100$ KΩ - 10 MΩ, while in the accumulation regime $|S|\approx500$ μV/K, $R_{sheet}\approx10$ KΩ and $n_{2D}\approx2\times10^{13}$ cm$^{-2}$. We note that the crossover between accumulation and depletion regimes may occur at very different threshold values of the gate voltage in different samples after the poling protocol, depending on the electrostatic landscape between the 2DEG and the back gate electrode, namely it is around zero voltage for the sample presented in the main text, around $V_g$=-20V for the sample in panel a), around $V_g$=+10V for the sample in panel b) and around $V_g$=+80V for the sample in panel c). For the latter sample, the threshold gate voltage even shifts among different poling runs.

**Supplementary note 2: Remark on absence of oscillations related to localized levels in the $R_{sheet}(V_g)$ curves.**

The simulated resistance shows oscillations in correspondence of the Fermi level crossing the localized levels, as does the Seebeck curve. However, these oscillations are not observed experimentally. We first point out that, from the experimental point of view, we carried out the resistance measurements with the highest possible level of accuracy and reliability, trying different configurations. In particular, we measured several samples in the maximum possible range of $V_g$, compatible with input impedance of nanovoltmeters, $R_{sheet}$ measurements were carried out using both d.c. (either applying alternating polarity or fixed polarity) and a.c. (lock in) techniques, we measured both voltage drops with a current bias and flowing current with a voltage bias, we took care of avoiding pinch off condition, we changed measuring parameters in such a way that the voltage drop in the transport measurement was similar to or even smaller than the voltage drop in the thermopower measurement, we measured data points closely spaced in the $V_g$ axis. In all our attempts, oscillatory behavior was never observed in the resistance curves of any of the investigated samples. We rule out the slight intrinsic spatial inhomogeneity in these systems as a cause for the absence of oscillations in $R_{sheet}$ curves, as it should have a smearing effect on $S$ and $R_{sheet}$ curves alike. On the other hand, similar discrepancy between $R_{sheet}(V_g)$ and $S(V_g)$ curves is not noticed here for the first time. In other systems, e.g. GaAs/AlGaAs [1,2] and MoS$_2$ [3], oscillations and divergence of $S$ at negative $V_g$ were observed, without corresponding oscillations in $R$. The key difference between Seebeck and resistance lies in the respective nature of these two measurements, namely the former is an open-circuit measurement with no net flow of electric charge, where the effect of the applied thermal gradient is counterbalanced by a voltage drop, while the latter is a closed-circuit measurement with injected current. We suggest that these differences, combined with strong non ohmic behavior of LAO/STO in the depletion regime, widely observed in literature [4], play a decisive role in washing out features associated to localized level filling in the resistance measurements. Similar non linear effects may be also responsible for the absence of oscillations in several resistivity measurements reported in literature [1-3], where diverging $R$ and $S$ in the depletion regime are likely associated to non ohmic transport mechanisms. On the contrary, literature works reporting oscillations of $S$ versus $V_g$, associated to very low and non-diverging resistance versus $V_g$ curves, evidence oscillations both in Seebeck and resistance curves [5,6].

Oppositely, the measurement of $S$ is unobtrusive and permits an ultra-refined scan of the unperturbed, closely-spaced polaronic densities lying below the band edge. Hence, the oscillations of $S$ should be thus ascribed to a paramount advantage of thermoelectric spectroscopy: the capability to probe extremely small changes in the charge density, without introducing significant perturbations. For this aspect, thermopower is far superior not only to resistivity measurements, but also to most of nowadays available spectroscopic techniques.

## Supplementary note 3: On the sensitivity of Seebeck measurement with respect to other spectroscopies

In the main article we claim that our Seebeck measurement is the first direct evidence of 'nearly-localized' states lying below the conduction band bottom. We need to specify the exact meaning of this strong statement, distinguishing what for us is 'direct' and 'indirect'.

Indirect evidences of charge localization under negative gate voltage are in fact abundant in literature, ranging from capacitance enhancement and negative compressibility [7] to metal-insulator transition at a specific charge threshold, magnetic correlations [8], and phase separation [9]. Furthermore, there is a number of spectroscopic measurements (see e.g. HAXPES spectra in [10]) where a peak of $Ti^{3+}$ character is clearly detected, attributed to localized electrons occupying a small fraction of Ti sites. All these evidences consistently suggest the presence of localization in some form in the highly depleted regime (typically below few $10^{12}$ cm$^{-2}$). On the other hand, what we define a 'direct' evidence has to do with a detailed scanning of the electronic structure of these states. This aspect has been, to our knowledge, lacking so far, and in fact, we claim that the Seebeck oscillations measured in this work are the first direct evidence of this sort. There are two major reasons at the basis of this unprecedented achievement:

a) the specific characteristics of the thermoelectric measurement, sensitive to charge density fluctuations as small as $10^{10}$ cm$^{-2}$ which translates, in terms of energies, to a resolution of the order of meV and below. This is a resolution that few spectroscopies can achieve, in practice.

b) the induction of a strongly-depleted regime: as long as the metallic regime is retained, no technique could be capable to distinguish the presence of tiny densities of polaronic levels. The *complete* depletion of the conduction band is 'conditio-sine-qua-non' for the emergence of localized behavior in transport, thermoelectric, or even electric and optical measurements.

Consider for example ARPES measurements, which is the prototypical approach to explore the electronic structure of a system: experiments carried out so far on LAO/STO have employed soft-X ray ARPES which at its best is performed with a resolution of 50 meV [11,12,13], i.e. a resolution way too low to distinguish levels separated a few meV one from each other and from the band bottom, as those described in our work. Furthermore, the well known "waterfall effects" creating blurred trails below the band bottom would make any attempt to distinguish features from these localized states totally hopeless. Finally, we point out that UV-ARPES with less than 10 meV resolution could be done, but the electron escape depth would be as short as few Angstrom, thus not enough to overcome the LAO barrier. More generally for what concern photoemission measurements, a long list of works can be found for LAO/STO with different techniques (HAXPES, RIXS, XPS, XAS, etc.) but after a careful exploration of this massive literature, we can safely conclude that none of those present the suited characteristics in terms of energy and charge density resolution to reveal the electronic structure at the same level of detail as that inferred from our Seebeck measurements.

A possible suitable technique to detect such localized states, could be scanning tunneling spectroscopy. With this technique, operating at low temperature (5K) it could be possible in principle to resolve electronic levels that are even more closely spaced than the ones observed in our experiment. However the design of such an experiment should face several difficulties, mostly related to the buried interface and to the contributions to the DOS coming also from surface and defect states. To the best of our knowledge, only one paper in literature shows scanning tunnel spectroscopy on LAO/STO but, in that case, data were averaged on 75 meV [14], so that features on the meV scale were not visible.

**Supplementary note 4: Phonon-drag model**

*Introduction* - Starting from the general expression for phonon-drag [15,16,17,18,19,20], we develop a specific formulation adapted to our means. We use a formalism based on 3D electronic structure coupled with 3D phonons, instead of the more common 2D electronic structure coupled with 3D phonons. Use of 3D formulas for the electrons is motivated by the need to treat on equal footing the bands and localized states, without a-priori assumption on the dimensionality. Of course, the electronic structure is fully anisotropic, and the 2D limit is easily recovered by imposing large effective masses in the interface-orthogonal direction. The multi-band effective mass modeling includes both delocalized conduction states and localized polaronic states (the latter having renormalized mass [21] and finite bandwidth) as shown in Figure 3 of the main text.

Since we are only interested in the low-T limit, only the interaction of electrons with acoustic phonons will be considered as a source of phonon-drag. The interaction is treated according to the deformation potential approach plus piezoelectric scattering to account for the polar character of the system. For simplicity, only one acoustic phonon branch and intra-band electron-phonon scattering is considered. These approximations may be seriously detrimental for detailed quantitative predictions, but are sufficient for the purpose of reproducing the order of magnitude and major features of the observed behavior.

From the coupled Boltzmann equations for electrons and phonons, the phonon-drag in the *j* direction is expressed as:

$$S_j^{\mathrm{g}} = \left( \frac{2e}{\sigma_j V K_{\mathrm{B}} T^2} \right) \sum_{n\mathbf{k}, n\mathbf{k'}, \mathbf{q}} \hbar \omega_{\mathbf{q}} \left( \frac{\Gamma_{n\mathbf{k}, n\mathbf{k'}}(\mathbf{q})}{\tau_{\mathrm{ph}}^{-1}(\mathbf{q}) + \tau_{\mathrm{ep}}^{-1}(\mathbf{q})} \right) V_j(n\mathbf{k}, n\mathbf{k'}, \mathbf{q}) \qquad (1)$$

where the factor 2 accounts for spin degeneracy, $e$ is the electron charge, $\sigma_j$ the electron conductivity, and $V$ the volume; $\mathbf{q}$ and $\omega_\mathbf{q}$ are phonon wavevector and frequency, respectively; $\Gamma_{n\mathbf{k},n\mathbf{k'}}(\mathbf{q})$ is the electron-phonon scattering rate (EPR) in k-space, $\tau_{ep}(\mathbf{q})$ the electron-phonon relaxation time, whereas $\tau_{ph}(\mathbf{q})$ includes all the other relevant phonon scattering processes (phonon-phonon scattering, boundary scattering, impurity scattering, etc.). $V_j$ is a velocity factor:

$$V_j(n\mathbf{k},n\mathbf{k'},\mathbf{q}) = v_j(\mathbf{q})\big[\tau(\mathbf{k})v_j(n\mathbf{k}) - \tau(\mathbf{k'})v_j(n\mathbf{k'})\big] \qquad (2)$$

where $\tau(\mathbf{k})$ is the electronic relaxation time, $v_j(\mathbf{q})$ and $v_j(n\mathbf{k})$ phonon and electron velocities, respectively. From Equation (1) it is apparent that phonon-drag essentially depends on two dominant factors: the inverse electric conductivity and the ratio of electron-phonon to total phonon scattering. Clearly, phonon-drag only matters if EPR is non-discardable with respect to the other phonon scattering processes. Following Cantrell notations [16], the inverse electron-phonon relaxation time (i.e. electron-phonon scattering frequency) is:

$$\tau_{ep}^{-1}(\mathbf{q}) = -\left(\frac{2}{N_\mathbf{q}^{'}K_B T}\right)\sum_{n\mathbf{k},n\mathbf{k'}}\Gamma_{n\mathbf{k},n\mathbf{k'}}(\mathbf{q}) \qquad (3)$$

$$\Gamma_{n\mathbf{k},n\mathbf{k'}}(\mathbf{q}) = f_{n\mathbf{k}}\left(1 - f_{n\mathbf{k'}}\right)N_\mathbf{q} A(\mathbf{q})\,\delta\big(\varepsilon_{n\mathbf{k'}} - \varepsilon_{n\mathbf{k}} - \hbar\omega_\mathbf{q}\big)\delta_{\mathbf{k'},\mathbf{k+q}} \qquad (4)$$

where $f_{n\mathbf{k}}$ and $N_\mathbf{q}$ are Fermi-Dirac and Phonon equilibrium distributions, respectively, $A(\mathbf{q})$ the electron-acoustic phonon coupling amplitude, and the two delta functions impose energy and momentum conservation for the absorption process (emission is accounted by the velocity factor). Equation (1) implicitly assumes the relaxation time approximation for phonons: the decrease of phonon population $N_\mathbf{q}^{Ph}$ due to phonon scattering is:

$$\frac{\partial N_{\mathbf{q}}^{Ph}}{\partial t} \approx -\left(\frac{\partial N_{\mathbf{q}}}{\partial q}\right)\left(\frac{1}{\tau_{\text{ph}}(\mathbf{q})} + \frac{1}{\tau_{\text{ep}}(\mathbf{q})}\right) \tag{5}$$

*Velocity factor* - In our model:

$$v_j(\mathbf{q}) = v_s \hat{q}_j ; \qquad v_j(n\mathbf{k}) = \frac{\hbar k_j}{m_{nj}^*} \tag{6}$$

where $v_s$ is the sound velocity, and $m_{nj}^*$ the $n^{\text{th}}$-band effective mass along j. From the linear-scaling approximation for the acoustic phonons we have $\omega_{\mathbf{q}} = \omega_q = v_s q$. Furthermore, we consider an isotropic phonon distribution, and put $q_j^2 = 1/3\ q^2$. Solving for $\delta_{k+q,k'}$ and assuming for simplicity $\tau(\mathbf{k}) = \tau(\mathbf{k}')$ we thus have:

$$V_j(n\mathbf{k}, n\mathbf{k}+\mathbf{q}, \mathbf{q}) = -\frac{v_s q_j}{q}\frac{\hbar q_j}{m_{nj}^*}\tau(\mathbf{k}) = -\frac{\hbar v_s q}{3 m_{nj}^*}\tau(\mathbf{k}) \tag{7}$$

The minus sign derives from the fact that for positive band curvature (thus electrons) the band velocity increases with $\mathbf{k}$.

*Energy-conserving delta function* - To handle non isotropic effective masses, it is useful to introduce the following change of variables:

$$k_j = K_j\sqrt{\frac{m_{nj}^*}{m}} ; \qquad q_j = Q_j\sqrt{\frac{m_{nj}^*}{m}} \tag{8}$$

where $m$ is an auxiliary mass. Thus, band energies can be rewritten at $2^{\text{nd}}$ order (setting the band bottom $\varepsilon_{n\mathbf{k}}^0 = 0$ for brevity):

$$\varepsilon_{n\mathbf{k}} = \frac{\hbar^2}{2}\left(\frac{k_x^2}{m_{nx}^*} + \frac{k_y^2}{m_{ny}^*} + \frac{k_z^2}{m_{nz}^*}\right) = \frac{\hbar^2 K^2}{2m} = \varepsilon_{nK} \qquad (9)$$

$$\varepsilon_{n\mathbf{k+q}} = \frac{\hbar^2(\mathbf{K+Q})^2}{2m} = \frac{\hbar^2 K^2}{2m} + \frac{\hbar^2 Q^2}{2m} + \frac{\hbar^2 KQ\cos\theta}{m} = \varepsilon_{nK+Q} \qquad (10)$$

where $\theta$ is the angle formed by vectors $\boldsymbol{K}$ and $\boldsymbol{Q}$. We then operate another change of variable from $\theta$ to $X$ defined as:

$$X = \frac{\hbar^2 K Q \cos\vartheta}{m} \qquad (11)$$

thus

$$\delta(\varepsilon_{n\mathbf{k+q}} - \varepsilon_{n\mathbf{k}} - \hbar\omega_q) = \delta(X - X_0) \qquad X_0 = \hbar\omega_q - \frac{\hbar^2 Q^2}{2m} \qquad (12)$$

The integral over $X$ is only non vanishing for $X_0$ included in the integral limits $X_{min} = -\hbar^2 K Q / m$ and $X_{max} = \hbar^2 K Q / m$, that is:

$$-\frac{\hbar^2 K Q}{m} \le \hbar\omega_q - \frac{\hbar^2 Q^2}{2m} \le \frac{\hbar^2 K Q}{m} \qquad (13)$$

the inequality can be rewritten:

$$\frac{\hbar^2(K-Q)^2}{2m} \le \varepsilon_{nk} + \hbar\omega_q \le \frac{\hbar^2(K+Q)^2}{2m} \qquad (14)$$

which has a simple interpretation: after absorption the carrier energy must be higher (lower) than the band energy corresponding to antiparallel (parallel) $\boldsymbol{K}$ and $\boldsymbol{Q}$ orientation. For conduction (valence) electrons, only the right (left) inequality matters.

*Sum over crystal momentum* - The sum over $\mathbf{k}$ is transformed first into a 3D-integral, then changed to an integral over $\boldsymbol{K}$ in polar coordinates. We integrate over the equatorial angle of $\boldsymbol{K}$ space, and take the azimuth angle as the θ between $\boldsymbol{K}$ and $\boldsymbol{Q}$, and then change θ to X. Solving for the energy delta function we finally obtain:

$$S_j^g = -\left(\frac{2e\, v_s^2}{3(2\pi)^2\, \sigma_j \mathrm{k_B T}^2}\right) \sum_{n=1}^{N_b} \frac{1}{m_{nj}^*} \sqrt{\frac{m_{nx}^* m_{ny}^* m_{nz}^*}{m}} \int dK\, K\, f_{nK}\, \tau(\mathbf{k})$$

$$\times \sum_{\mathbf{q}} \frac{1}{Q} \frac{q^2 N_q A(\mathbf{q})}{\tau_{\mathrm{ph}}^{-1}(\mathbf{q}) + \tau_{\mathrm{ep}}^{-1}(\mathbf{q})} \left(1 - f_{nK+Q}\right) \tag{15}$$

Here $N_b$ is the number of bands. The minus sing comes from the velocity factor. Finally, $K$ can be changed with the energy (Equation (10)):

$$S_j^g = -\left(\frac{2e\, v_s^2}{3(2\pi)^2\, \hbar^2 \sigma_j \mathrm{k_B T}^2}\right) \sum_{n=1}^{N_b} \frac{m}{m_{nj}^*} \sqrt{\frac{m_{nx}^* m_{ny}^* m_{nz}^*}{m}} \int_{\varepsilon_n^0}^{\varepsilon_n^0+W} d\varepsilon\, f_\varepsilon\, \tau(\varepsilon)$$

$$\times \sum_{\mathbf{q}} \frac{1}{Q} \frac{q^2 N_q A(\mathbf{q})}{\tau_{\mathrm{ph}}^{-1}(\mathbf{q}) + \tau_{\mathrm{ep}}^{-1}(\mathbf{q})} \left(1 - f_{\varepsilon+\hbar\omega_q}\right) \tag{16}$$

where $\varepsilon_{nk}^0$ and $W_n$ are band bottom and bandwidth, respectively, and we have assumed, as customary, that the electronic relaxation time only depends on the wavevector through the energy.

*Integration over phonon wavevector* - The modulus of the rescaled variable $\boldsymbol{Q}$ in Equation (16) includes the band masses according to Equation (8); this would require to solve a 3D integration over the phonon wavevector. To simplify the calculation, we assume the following approximation:

$$Q \approx \sqrt{\frac{m}{\tilde{m}_n}}\, q; \qquad \tilde{m}_n \approx \frac{m_{nx}^* m_{ny}^* m_{nz}^*}{m_{nx}^* m_{ny}^* + m_{nx}^* m_{nz}^* + m_{ny}^* m_{nz}^*} \tag{17}$$

where $\tilde{m}_n$ is the geometrically averaged effective mass. As showed later on, scattering amplitude $A$ and phonon relaxation times $\tau_{\mathrm{ph}}$ and $\tau_{\mathrm{ep}}$ actually depend only on the modulus $q$; thus, we can easily transform the sum over $\mathbf{q}$ in Equation (16) in the integral, and solve it in polar coordinates:

$$S_j^g = -\left(\frac{V e v_s^2}{12\pi^4 \hbar^2 \sigma_j k_B T^2}\right) \sum_{n=1}^{N_b} \tilde{m}_{nj} \int_{\varepsilon_n^0}^{\varepsilon_n^0+W} d\varepsilon\, \tau(\varepsilon) f_\varepsilon$$

$$\times \int_0^{q_D} dq \left(\frac{q^3 N_q A(q)}{\tau_{ph}^{-1}(q) + \tau_{ep}^{-1}(q)}\right)\left(1 - f_{\varepsilon+\hbar\omega_q}\right) \tag{18}$$

Here the phonon wavevector is integrated up to the Debye frequency $q_D = k_B T_D / v_s \hbar$, where $T_D$ is the Debye temperature (~500 K for STO [22]), and:

$$\tilde{m}_{nj} = \frac{1}{m_{nj}^*} \frac{m_{nx}^* m_{ny}^* m_{nz}^*}{\sqrt{m_{nx}^* m_{ny}^* + m_{nx}^* m_{nz}^* + m_{ny}^* m_{nz}^*}} \tag{19}$$

*Electron-phonon coupling* - In the simplest treatment, the electron-acoustic phonon scattering for an isotropic phonon distribution can be taken as the sum of two terms [23,24]:

$$A(q) = \left(\frac{\pi D^2}{V\rho\, v_s}\right) q + \left(\frac{\pi e^2 v_s K_{em}^2}{V \kappa_0 \kappa}\right) \frac{q^3}{\left(q^2 + q_0^2\right)^2} \tag{20}$$

where $D$ is the deformation potential, $\rho$ the mass density, $\kappa_0$ and $\kappa$ vacuum permittivity and dielectric constant, $K_{em}$ the 3D-averaged electromechanical coupling, and $q_0$ the Debye screening length. The first term ("deformation potential scattering") describes the coupling of electrons with long-wavelength longitudinal acoustic waves treated as an homogeneous strain. The second ("piezoelectric scattering") is the additional contribution due to the coupling with the electric field produced by the strain. For a non-polar system ($K_{em} = 0$) or in the limit of large doping concentration (i.e. strong Debye screening) the second term vanishes and only the deformation potential contributes to the acoustic scattering. In case of small screening ($q_0 = 0$) and highly ionic compounds, on the other hand, the piezoelectric contribution (~1/$q$) may become dominant at small $q$. For some parameters appearing in Equation (22) the appropriate values to be used for our LAO/STO model are not easily determined, since these measurements may crucially depend on structural and chemical composition of samples, as well as doping type and concentration, thus they should be considered as merely indicative. We take $K_{em} = 0.37$, i.e. the average of planar and transversal components measured for a solid solution of Ti-based ceramics [25]. The dielectric constant under large negative gate is also difficult to be quantified reliably. It has been demonstrated [26] that below $T$=50 K a large negative gate field induces a polar transition in STO which sweeps away the quantum paraelectric state and largely reduces the screening capability. Tentatively, we use $k = 300$, that is the room-T value for STO bulk. Furthermore, we can assume a very long Debye length so that carrier screening is discardable ($q_0 \approx 0$). This is indeed a basic hypothesis for having strong electron-phonon scattering [27], i.e. in the low-density charge-localized limit the carrier mobility is so small that the piezoelectric interaction becomes unscreened. Finally for LAO/STO we use D=8.2 eV (used for GaAs in Ref. [16]) and $v_s = 0.05\,\mathrm{m\,sec^{-1}}$ (that is the 3D-average sound speed at low temperature [28]).Equation (22) is then inserted into Equation (20) to obtain:

$$S_j^g = -\left( \frac{e\, v_s^2}{12\pi^3\hbar^2\sigma_j k_B T^2} \right) \sum_{n=1}^{N_b} \tilde{m}_{nj} \int_{\varepsilon_n^0}^{\varepsilon_n^0+W} d\varepsilon\, \tau(\varepsilon) f_\varepsilon$$

$$\times \int_0^{q_D} dq \left( \frac{N_q}{\tau_{ph}^{-1}(q) + \tau_{ep}^{-1}(q)} \right) \left( C_{DP}\, q^4 + C_{PZ}\, \frac{q^6}{(q^2+q_0^2)^2} \right) \left( 1 - f_{\varepsilon+\hbar\omega_q} \right) \qquad (21)$$

$$C_{DP} = \frac{D^2}{\rho\, v_s}; \qquad\qquad C_{PZ} = \frac{e^2\, v_s\, K_{em}}{\kappa_0 \kappa} \qquad\qquad (22)$$

which is finally integrated numerically.

## Supplementary note 5: Phonon relaxation time

For the electron-acoustic phonon scattering we start from:

$$\tau_{ep}^{-1}(\mathbf{q}) = -\left(\frac{2}{N_\mathbf{q}' K_B T}\right) \sum_{n\mathbf{k},n\mathbf{k}'} \Gamma_{n\mathbf{k},n\mathbf{k}'}(\mathbf{q})$$

$$= -\left(\frac{2N_\mathbf{q} A(\mathbf{q})}{N_\mathbf{q}' K_B T}\right) \sum_{n\mathbf{k}} f_{n\mathbf{k}}\left(1 - f_{n\mathbf{k}+\mathbf{q}}\right) \delta\left(\varepsilon_{n\mathbf{k}+\mathbf{q}} - \varepsilon_{n\mathbf{k}} - \hbar\omega_\mathbf{q}\right) \qquad (23)$$

We can proceed as before, applying ellipsoidal band approximation for the electrons and linear-scale approximation for acoustic phonons, then changing the sum over $\mathbf{k}$ with the integral in polar coordinates; the energy delta function thus depends only on the modulus of $\mathbf{K}$, $\mathbf{Q}$ and their angle $\theta$, and we can exploit the same variable substitutions to obtain:

$$\sum_{n\mathbf{k}} f_{n\mathbf{k}}\left(1 - f_{n\mathbf{k}+\mathbf{q}}\right) \delta\left(\varepsilon_{n\mathbf{k}+\mathbf{q}} - \varepsilon_{n\mathbf{k}} - \hbar\omega_\mathbf{q}\right) = \frac{V}{(2\pi)^2 \hbar^4 q} \sum_n \tilde{m}_n^2 \int_{\varepsilon_n^0}^{\varepsilon_n^0+W} d\varepsilon \, f_\varepsilon\left(1 - f_{\varepsilon+\hbar\omega_q}\right) \quad (24)$$

where:

$$\tilde{m}_n^2 = \frac{m_{nx}^* m_{ny}^* m_{nz}^*}{\sqrt{m_{nx}^* m_{ny}^* + m_{nx}^* m_{nz}^* + m_{ny}^* m_{nz}^*}} \qquad (25)$$

furthermore:

$$\frac{N_q}{N_q'} = -\left(\frac{K_B T}{1 + N_q}\right) \qquad (26)$$

thus, at the end:

$$\tau_{ep}^{-1}(q) = \left(\frac{V A(q)}{2\pi^2 \hbar^4 q(1 + N_q)}\right) \sum_n \tilde{m}_n^2 \int_{\varepsilon_n^0}^{\varepsilon_n^0+W} d\varepsilon \, f_\varepsilon\left(1 - f_{\varepsilon+\hbar\omega_q}\right) \qquad (27)$$

$$= \left(\frac{C_{DP}}{2\pi\hbar^4} + \frac{C_{PZ} q^2}{2\pi\hbar^4 \left(q^2 + q_0^2\right)^2}\right) \frac{1}{(1 + N_q)} \sum_n \tilde{m}_n^2 \int_{\varepsilon_n^0}^{\varepsilon_n^0+W} d\varepsilon \, f_\varepsilon\left(1 - f_{\varepsilon+\hbar\omega_q}\right) \qquad (28)$$

For what concerns other phonon scattering processes, a simple parametrization is proposed in the seminal article of Callaway [29]:

$$\tau_{\mathrm{ph}}^{-1}(q) = A\,\omega_q^4 + BT^3\omega_q^2 + \frac{v_s}{L} \qquad (29)$$

where A describes the scattering by point impurities, B the phonon-phonon scattering including both normal (crystal momentum-conserving) and umpklapp scattering, and $v_s/L$ the boundary scattering ($L$ is a characteristic sample length; in our model $L=1$ mm). In the following numerical examples we will use values for A and B given in Ref. [29]. While probably not appropriate for quantitative predictions, Equation (29) is sufficient for the purpose of qualitative description of general trends.

**Supplementary note 6: Diffusive Seebeck model**

The formulation of diffusive Seebeck for a multiband effective-mass model is based on the Boltzmann Transport Equation in relaxation time approximation, and band energies modeled by non-isotropic effective masses. The model is described in detail in Ref. [30] where it was first applied to LAO/STO. Here we can just recall the final expression. The DC conductivity (in 3D) in direction $j$ is given by:

$$\sigma_j = \frac{\sqrt{2}\,e^2}{\pi^2\hbar^3 k_B T}\sum_{n=1}^{N_b}\frac{\sqrt{m_{nx}^* m_{ny}^* m_{nz}^*}}{m_{nj}^*}\int_{\varepsilon_n^0}^{\varepsilon_n^0 + W_n} d\varepsilon\,\tau(\varepsilon)\left(-\frac{\partial f}{\partial \varepsilon}\right)\left(\varepsilon - \varepsilon_n^0\right)^{3/2} \qquad (30)$$

The Seebeck associated to the $n^{\mathrm{th}}$ band is $S_{nj} = \Lambda_{nj}/\sigma_j$, where:

$$\Lambda_{nj} = -\frac{\sqrt{2}\,e N_b}{\pi^2\hbar^3 k_B T^2}\frac{\sqrt{m_{nx}^* m_{ny}^* m_{nz}^*}}{m_{nj}^*}\int_{\varepsilon_n^0}^{\varepsilon_n^0 + W_n} d\varepsilon\,\tau(\varepsilon)\left(-\frac{\partial f}{\partial \varepsilon}\right)(\varepsilon - \varepsilon_F)\left(\varepsilon - \varepsilon_n^0\right)^{3/2} \qquad (31)$$

Finally, the diffusive Seebeck is obtained as:

$$S_j = \left(\frac{1}{N_b}\right)\sum_{n=1}^{N_b} S_{nj} = \left(\frac{1}{\sigma_j N_b}\right)\sum_{n=1}^{N_b}\Lambda_{nj} \qquad (32)$$

Notice that, although explicitly dependent on the inverse conductivity, $S_j$ depends smoothly on the fundamental ingredients which determine the conductivity amplitude (relaxation time and electron velocities) since they equally appear in the numerator and denominator in $S_{nj} = \Lambda_{nj}/\sigma_j$. At low $T$ the Fermi function derivative reduces to a delta function, and the integrands of $\Lambda_{nj}$ and $\sigma_j$ almost cancel out. It follows that in practice, it is not possible to obtain a diverging diffusive Seebeck as a consequence of the conductivity suppression. This represents an intrinsic threshold to the maximum amplitude that can be reached by diffusive Seebeck (according to the literature, this should be of the order of $10^3\,\mu V/K$). On the other hand, the $S_j^g$ dependence on the inverse conductivity (Equation (21)) is real and does not cancel at low temperature, thus no such limitation exists on the attainable amplitude.

## Supplementary note 7: Electronic relaxation time

Since we are mainly interested in the very low-$T$ regime, we can consider a minimal model for the electronic relaxation time including acoustic-phonon scattering (AP) and impurity scattering (IS). AP is treated within elastic deformation potential approach:

$$\tau_{AP}^{-1}(\varepsilon) = \frac{(2\tilde{m})^{3/2} K_B T D^2 \varepsilon^{1/2}}{2\pi\hbar^4 \rho\, v_s^2} \tag{33}$$

where the electron energy is relative to the CBB. For IS we adopt the well known Brooks-Herring formula:

$$\tau_{IS}^{-1}(\varepsilon) = \frac{\pi n_I Z^2 e^4 \varepsilon^{-3/2}}{\sqrt{2\tilde{m}}(4\pi\kappa_0\kappa)^2}\left[\log\left(1 + \frac{8\tilde{m}\varepsilon}{\hbar^2 q_0^2}\right) - \frac{1}{1 + \left(\hbar^2 q_0^2 / 8\tilde{m}\varepsilon\right)}\right] \tag{34}$$

where $n_I$ is the impurity concentration, $Z$ the impurity charge. For localized states, we assume the following hopping frequency expression:

$$\tau_{\text{h}}(\varepsilon) = \tau_{\text{h0}} \, e^{-\left(\frac{\varepsilon_{\text{CBB}}-\varepsilon}{K_B T}\right)} + \tau_{\text{h0}} \, e^{-\left(\frac{E_{\text{T}}}{K_B T}\right)^{1/3}} \qquad (35)$$

where $\tau_{\text{h0}}$ is a characteristic hopping time, and the two exponentials represent the probability of hopping by thermal excitation (for energies lower than the mobility edge $\varepsilon_{\text{CBB}}$) and by tunneling across an energy barrier $E_{\text{T}}$, respectively. For the LAO/STO well simulation we use $\tau_{\text{h0}} = 10^{-11}$ sec, and $E_{\text{T}} = 3$ meV (that is in our model the energy separation between two localized states).

**Supplementary note 8: Test cases**

*Bulk SrTiO₃ n-doped (band regime)* - In its simplest form, the electronic structure of n-doped bulk STO can be configured in terms of three $t_{2g}$ degenerate conduction bands of $d_{xy}$, $d_{xz}$, $d_{yz}$ character with strongly anisotropic effective masses: $m^{*}_{xy,j} = (0.7, 0.7, 8.8)m_{\text{e}}$, $m^{*}_{xz,j} = (0.7, 8.8, 0.7)m_{\text{e}}$, and $m^{*}_{yz,j} = (8.8, 0.7, 0.7)m_{\text{e}}$. Results for $n_{\text{3D}} = 2 \times 10^{19}$ cm⁻³ are shown in Supplementary Figure 2. In this example the electron-phonon scattering frequency $\tau_{\text{ep}}^{-1}$ (Supplementary Figure 2(A)) only includes the deformation potential (i.e. $K_{\text{em}}=0$). We see that $\tau_{\text{ep}}^{-1}$ grows with the phonon frequency at fixed $T$, and decreases at a given phonon frequency with increasing T. Its dependence on $\hbar\omega_q$ and $T$ is dominated by two factors: the inverse Bose occupancy, and the integral over the electron-hole occupancy (see Equations (28), (29) and (30)). For what concern the $\tau_{\text{ep}}^{-1}$ dependence on the integral, in the $T=0$ limit the phonon can only be absorbed by electrons in the energy range $\varepsilon_n \subseteq [\varepsilon_{\text{F}} - \hbar\omega_q, \varepsilon_{\text{F}}]$. Thus, the number of available electronic states, and so the scattering rate, grows linearly with $\hbar\omega_q$ and saturates at $\hbar\omega_q = (\varepsilon_{\text{F}} - \varepsilon_0)$ corresponding to the full occupied range $\varepsilon_n \subseteq [\varepsilon_0, \varepsilon_{\text{F}}]$. With increasing $T$ this condition is progressively relieved by the partial occupancy, saturation occurs at much higher values and it is not sharp anymore. At fixed $\omega_q$, on the other hand, the occupancy factor $f_\varepsilon \left(1 - f_{\varepsilon+\hbar\omega_q}\right)$ decreases with increasing $T$.

In Supplementary Figure 2(B) we report the total phonon scattering frequency $\tau_{ep}^{-1} + \tau_{ph}^{-1}$. Since $\tau_{ph}^{-1}$ has a strong $\sim T^3$ dependence, it becomes preponderant over electron-phonon at increasing $T$, causing the phonon-drag to quickly fade away with increasing $T$. Indeed, only at a very low temperature the profile of total scattering is visibly affected by electron-phonon. The phonon-drag $T$-behavior can be understood considering the relative scattering ratio (see Equation (1) and (3)):

$$ S^{g} \approx \frac{N_{q}^{'}}{k_B T}\left( \frac{\tau_{ep}^{-1}(q)}{\tau_{ph}^{-1}(q) + \tau_{ep}^{-1}(q)} \right) \qquad (36) $$

displayed in Supplementary Figure 2(C); starting from low $T$, we see that the ratio first quickly grows with $T$, reaching its maximum at about 50-60 K ($\sim T_D/10$, where $T_D$=500 K), and then it smoothly decreases. In Supplementary Figure 2(E) the phonon-drag is shown, together with diffusive and total Seebeck; consistently with Supplementary Figure 2(C), phonon-drag reaches its maximum magnitude at about 50 K and vanishes at room $T$. We stress that our phonon-drag modeling is only based on acoustic-phonon scattering, while at room $T$ other electron-phonon scatterings (primarily polar optical phonon scattering) are expected to dominate. Thus albeit qualitatively reasonable, these results have the only purpose of illustrating the model characteristics.

*Bulk SrTiO₃ n-doped (band plus localized states)* - It is well known that the transport properties of STO are characterized by polaronic behavior and electron localization. Similar behavior has been also reported for LAO/STO [31]. A simple way to model these states is through poorly-dispersed mass-renormalized bands laying below the CBB, and assuming that conduction across these states can only occur by hopping. In the example shown in Supplementary Figure 3 we consider three $t_{2g}$ conduction bands (as in the previous test-case) plus one localized state centered 5 meV below the CBB, with W=10 meV and $m_{loc,j}^{*}$= (3.0, 3.0, 8.8)$m_e$. For the same charge density ($n_{3D}$= 2×10$^{19}$ cm$^{-3}$) the electron-phonon scattering at low phonon frequency (Supplementary Figure 3(A)) is enhanced with respect to the previous case (Supplementary Figure 2(A)) and shows a dome-like feature of width equal to $W$. The dome is progressively smoothed with increasing $T$. In the total scattering frequency (Supplementary Figure 3(A)) the electron-phonon dome is well visible up to about $T$=50 K, while above $T$=50 K it becomes more and more inessential (we use blue to red color change to emphasize the transition).

The localized state presence thus rises the phonon drag amplitude and shifts its maximum weight towards lower temperatures: phonon drag (Supplementary Figure 3(D), violet circles curve) is now peaked at about 40 K and visibly enhanced in amplitude (140 μV/K) with respect to the previous case (100 μV/K). We also monitored the phonon-drag change with charge density. For increasing density, the system behavior becomes progressively band-like and phonon-drag is reduced accordingly. The system evolution with density can be understood from resistivity (Supplementary Figure 3(C)): for $n_{3D}$= 2×10^{19} cm^{-3} a temperature-driven metal-insulating transition is visible, due to the presence of the localized state which at $T$=0 hosts all the available charge (in the previous test case, the same charge density was fully hosted by the conduction bands, and the resistivity was metallic-like). For increasing density, the zero-$T$ Fermi energy progressively rises, until it is brought above the CBB, whence resistivity becomes metallic-like through the whole temperature range.

*LAO/STO well* - The model is built to reproduce as closely as possible the most important features of the Seebeck measured under negative voltage at $T$=4.2 K (we are specifically referring to the LAO/STO sample displayed in the main text): a) the distance in $V_g$ between two consecutive oscillations (about 0.6 V); b) the charge density change corresponding to each oscillation (about 10^{10} electrons × cm^{-2}); c) the diverging values of $S$, increasing in amplitude by about 3 orders of magnitude across a $V_g$ interval of about 14 V, with 12 oscillations interpreted as the crossing of $E_F$ through an equal number of polaronic levels. All these required features represent sharp constraint on the basic characteristics of the model (energies, effective masses, bandwidth), with little freedom left for alternative arrangements. Very importantly, as explained in the main text, these features in no case can be matched with the diffusive Seebeck regime, but can be only accounted by phonon-drag.

In order to reproduce the 12 oscillations displayed by the measurement on the main sample, we include in the model a series of 12 localized states, regularly distributed in energy below the CBB (parameters used for the model are listed in Supplementary Table 1; charge density and density of states of this electronic structure are displayed in Figure 3 of the main article). These localized states, separated from each other by 3 meV at the band bottom, are not identical: moving from higher to lower energies, the associated bandwidth and DOS become smaller and smaller, while the effective masses become larger, so that charge localization is progressively enhanced while $E_F$ moves away from the CBB.

For what concern the fundamental understanding of the localized states, in literature they are typically related to two possible sources, which however do not exclude each other and can occur simultaneously: a) Anderson localization (disorder) which is known to affect the LAO/STO interface; (the presence of an Anderson tail of localized states just below the conduction edge of oxide heterostructures is well documented in literature [32]); b) Mott localization, invoked to explain other correlation phenomena observed in this system (see e.g. capacitance enhancement described in [7]). Indeed, at very low charge density, the charge can localize in $Ti^{3+}$ sites, eventually helped by structural deformations (polarons) even in absence of structural disorder.

To mimic the experiment, the simulation assumes that at zero voltage the lowest conduction band is occupied by a mobile carrier density ($n_{2D} \sim 1.2 \times 10^{13} cm^{-2}$) typical for LAO/STO. Then, it is assumed that the leading effect of a negative $V_g$ is the progressive charge depletion, thus mapping $S(E_F)$ is substantially equivalent to mapping $S(V_g)$. Seebeck and resistivity are then calculated as a function of the progressively decreasing $E_F$. Additional effects of negative $V_g$, such as a possible change in the

STO dielectric permittivity at the interface [26], are inessential since our simplified model only includes by construction electronic states confined in a single interface layer. We underline that specific *quantitative* features of the model (energies, effective masses, bandwidths) are set to reproduce as closely as possible the characteristics of the $S$ measurements in terms of number of oscillations, oscillation amplitude, $S$ vs. $V_g$ absolute value. Of course these characteristics change, within a certain extent, from sample to sample, and so must do the parameters of the model. In particular the values in Supplementary Table I chosen for the numerical simulation are adapted to the sample shown in the main part of the article.

The resulting phonon-drag, diffusive Seebeck, and resistivity are displayed in Figure 3 of the main text. Here we add (Supplementary Figure 4) the analysis of their state-by-state contributions, which is useful to highlight some key aspects of their behavior.

Consider first the diffusive Seebeck of each individual state (Supplementary Figure 4B)): each of them displays a negative-$S$ region followed, after crossing the $S=0$ axis, by a positive bell-like feature. Indeed, according to the Mott formula [33] $S_d \approx -[\partial n/\partial \varepsilon]_{\varepsilon=\varepsilon_F}$, $S_d$ must change from negative to positive values while moving through the localized DOS, corresponding to positive and negative DOS derivatives in the regions before and after the DOS peak. This captures a general feature of diffusive Seebeck: $S_d$ must necessarily oscillate through the zero while moving across localized DOS, where with 'localized' it is implied a series of narrow DOS, either separated by energy gaps or even partially overlapping, but with well distinct peaks (this is the case of our model). This constraint does not apply to the state-resolved $S_g$ (Supplementary Figure 4A)) which indeed do not change sign while $E_F$ moves across the localized states.

Notice also that for a model made of multiple overlapping bands shifted from each other in energy at the bottom, $E_F$ would not cross any DOS peaks, and in turn $S_d$ could display non-zero centered oscillations corresponding to the $E_F$ crossing the onset of each band bottom. However, as extensively discussed in the main article, in the band regime the amplitude of diffusive Seebeck at low $T$ is totally discardable on the scale of the measured Seebeck. In our intensive attempts to simulate the experiment on the basis of a multi-band modeling, $S_d$ could never be found to exceed a few tents of μV/K at low $T$, no matter how large the DOS slope (i.e. the effective masses) could be. On the basis of this analysis, we can confidently exclude that diffusive Seebeck could ever cause the huge, furiously oscillating thermopower observed in the measurements.

Finally, in Supplementary Figure 5 we report the electron-phonon scattering frequency relative to our localized state model, as a function of the phonon frequency at varying 2D doping concentration and fixed $T$=4.2 K. In Supplementary Figure 5(A) only the deformation potential contribution is included. The dome-like feature is similar to that seen in Supplementary Figure 3(A), but the values are much amplified (by order of magnitudes) for this extremely low-density regime. In Supplementary Figure 5(B) we show again the electron-phonon scattering frequency, but also adding the piezoelectric contribution. The latter becomes dominant in the very low frequency range, as a consequence of the $A(q) \approx q^{-1}$ behavior (Equation (22)). We see that starting from low doping (red solid curve), the scattering first increases with $E_F$ due to the progressive increase of $W$ and, in turn, of available electronic transitions. Then, a regime change occurs (highlighted by the red-to-blue color change) and the scattering start to decrease with $E_F$ in consequence of the progressive decrease of the effective masses. For each curve, the phonon frequency region where the scattering is effective roughly follows the DOS profile of the electronic state crossed by $E_F$ at that doping.

**Supplementary References**